\documentclass[12pt,preprint]{aastex}

\shorttitle{Putative Pulsar in SN1987A}
\shortauthors{\"Ogelman \& Alpar}

\begin{document}


\title{Constraints on a putative pulsar in SN1987A}


\author{H. \"{O}gelman\altaffilmark{1,2}} 
\email{ogelman@cow.physics.wisc.edu}

\author{M. A. Alpar\altaffilmark{2}}
\email{alpar@sabanciuniv.edu}


\altaffiltext{1}{Department of Physics , University of Wisconsin at Madison, 2531 Sterling Hall, Madison, WI 53706}
\altaffiltext{2}{Faculty of Engineering and Natural Sciences, Sabanc{\i} University, Orhanl{\i}-Tuzla, Istanbul 34956, Turkey}


\begin{abstract}
We assume that all the energy loss of the putative pulsar in SN 1987A would contribute to the luminosity of the remnant, which acts like a bolometer. The bolometric luminosity of SN 1987A provides an upper bound on the pulsar's rate of energy loss. An isolated pulsar spinning down by magnetic dipole radiation alone, with initial rotation periods of 10-30 ms, as extrapolated for galactic young pulsars, can have a luminosity below the bolometric bound if either the magnetic field is weak, B $\sim$ 10$^{9}$-10$^{10}$ G or if it is so strong that the pulsar luminosity decays rapidly, 
B $\sim$10$^{16}$ G.
\end{abstract}




\section{Introduction}

The observed Balmer absorption lines class SN 1987A as a Type II supernova. The identification 
of the pre-supernova star, the neutrino flux detected at the time of the explosion, the 
evolution of the optical light curve and the observations of the emission of soft $\gamma$ 
rays all confirm this classification. Therefore, it is expected that a neutron star lies at 
the center of the remnant. Simulations of the neutrino output in core collapse leading to 
a black hole differ strongly from the Kamiokande neutrino detections, which agree with 
several neutron star models (Burrows 1988). Any detection of a neutron star formed in 
SN1987A would arouse great interest since the properties of very young neutron stars 
are largely unknown. Searches for pulsed emission from the remnant have yielded no 
results (\"{O}gelman et al. 1990). A claimed pulsar period of 2.14 ms has been discussed by Middleditch et al (2000). 
A synchrotron nebula would indicate a young 
pulsar even if its beam does not sweep our direction so that a pulsed signal is 
not received. No synchrotron nebula has been detected in the remnant of SN 1987A, 
with a Chandra upper limit of 5.5 $\times$ 10$^{33}$ ergs s$^{-1}$ to the luminosity 
of a central X-ray source (Park et al. 2003). 

The newborn neutron star is characterized by its initial rotation rate and magnetic dipole 
moment. It will spin down, generating radiation and accelerating charged particles at the 
expense of its rotational energy. The power output of the neutron star is the luminosity 
of a rotating dipole, regardless of the detailed processes by which the dipole emission 
is converted to the energy of charged particle acceleration and high-enegy electromagnetic 
radiation. The remnant is optically thick to the radiation emitted by the young pulsar. 
Thus, the remnant will act as a bolometer, absorbing and reprocessing the pulsar luminosity. 
Most of the remnant's central luminosity is in the optical, IR and UV bands, exceeding 
its X-ray luminosity by several orders of magnitude.  The observed bolometric luminosity 
of the remnant is dominated and well fitted with the radioactive decay luminosity, so the 
contribution from the pulsar must be a small admixture. We shall assume that the pulsar loses 
energy as an isolated rotating dipole, of constant magnetic dipole moment, without any 
other torques (as would arise, for example, from the presence of a fallback disk). 
We use the bolometric luminosity to place bounds on the dipole emission of the 
putative pulsar and on its birth properties. 

\section{Estimation of the pulsar's birth properties}

The spindown of the pulsar by dipole radiation follows from
\begin{equation}
I \Omega \dot{\Omega} = L_{dip} = \frac{2 \mu^2 \Omega^4}{3 c^3}
\end{equation}
where I is the moment of inertia of the neutron star,  $\Omega$ is the star's rotation 
rate and $\mu$ is 
its magnetic moment. The solution for the luminosity is
\begin{equation}
L_{dip}(t) = L_{0} ( 1 + t / t_{0} )^{-2}.
\end{equation}
in terms of the initial luminosity L$_{0}$ = 2/3 $\mu^2 \Omega(0)^4 / c^3$ . 
The timescale t$_{0}$ is the initial energy loss timescale of dipole radiation, 
\begin{equation}
t_{0}= E_{0} / L_{0} = \frac{3  I c^3}{4 \mu^2 \Omega(0)^2} 
\end{equation}
where E$_{0}$ = 1/2 I $\Omega(0)^{2}$ is the initial rotational energy.

The bound provided by the observed bolometric luminosity at time t 
is $L_{dip}(t) <  L_{bol}(t)$. Using Eq. (2) and substituting for L$_{0}$ and t$_{0}$, 
we obtain
\begin{equation}
\left( 1 + \frac{4 \mu^2 \Omega(0)^2 t}{3  I c^3 } \right)^2 > \frac{ 2 \mu^2 \Omega(0)^4}{3 c^3 L_{bol}(t)} .
\end{equation}
Expressing the inequality in terms of the  initial period P$_0$ and 
magnetic moment $\mu$ and taking the square root leads to 
\begin{equation}
({P_0}^2 + \frac{16 \pi^2 \mu^2 t}{3  I c^3 }) > \frac{ 4 \pi^2 \mu }{ (3/2 c^3 L_{bol}(t))^{1/2}} .
\end{equation}
Completing the square in terms of $\mu$ and normalizing, 
we obtain the constraint 
\begin{equation}
\frac{{P_0}^2}{{\bar{P}(t)}^2} + \frac{(\mu - \bar{\mu}(t))^2}{{\bar{\mu}(t)}^2} > 1.
\end{equation}

Thus, the allowed values of the magnetic moment $\mu$ and initial 
period P$_0$ must lie outside an ellipse in the P$_0$-$\mu$ plane, with
\begin{eqnarray}
\bar{P}(t) &=& ( \frac{I \pi^2}{2 t L_{bol}(t)})^{1/2} \\
\bar{\mu}(t) &=& \frac{(6c^3)^{1/2} I }{8 t (L_{bol}(t))^{1/2}}
\end{eqnarray}
as shown in Fig. 1.  
For initial rotation periods P$_0 < \bar{P}$, the dipole luminosity is less than the 
observed L$_{bol}$(t), either because the magnetic moment is small or the magnetic 
moment is large enough that the dipole luminosity has decreased rapidly. 
For P$_0 > \bar{P}$, the initial rotation rate is slow enough that the pulsar's 
luminosity at time t is less than L$_{bol}$(t) for all values of the magnetic moment. 
 
As the observed bolometric luminosity decays rapidly with time, the forbidden ellipse 
in the P$_0$-$\mu$ plane expands, as indicated by Eqs. (5) and (6). Thus, in a 
data set L$_{bol}$(t)  extending to t = t$_{max}$ it is the lowest 
luminosity L$_{bol}$(t$_{max}$) that yields the tightest constraint, 
encompassing the constraints imposed by all earlier observations. The bolometric 
luminosity observed for over 10 yr is fitted quite well by a series of exponential 
decays corresponding to the radioactive decays of $^{56}$Co, $^{57}$Co, $^{44}$Ti, $^{22}$Na, 
and $^{60}$Co (Timmes et al. 1996). Bouchet et al. (1991) have compiled the bolometric 
luminosity evolution 
in the IR, optical and UV bands for days 14-432 after the supernova. 
An initial hump is followed after 130 days by an exponential decay of time 
constant $\tau \cong $110 days, corresponding to the decay of $^{56}$Co (see Fig. 3. of Bouchet et al. (1991)), 
with a slight increase in the rate 
of decay after t= 300 days. At later times the bolometric luminosity can be 
fitted with a sum of slower decays. At 3346 days the bolometric 
luminosity L$_{bol}$ = 2.1 $\times$ 10$^{36}$ ergs s$^{-1}$, fitting on a curve corresponding 
to the decay of $^{44}$Ti (Balberg et al. 1999). Later data taken by the Hubble Space 
Telescope Wide Field Planetary Camera and Wide Field Planetary Camera 2 give UBVRI magnitudes corrected by subtraction of the ring 
contribution in the SN1987A images (Soderberg, Challis \& Suntzeff 1999). We use the 
latest observations reported 
in this work, deriving a bolometric luminosity L$_{bol}$ = 3 $\times$ 10$^{34}$ ergs s$^{-1}$ 
from the UBVRI magnitudes at 4339 days, using a distance of 55 kpc for the LMC. The 
magnitudes were converted to luminosities using the standard bandwidths and conversion formulae
for the UBVRI bands (Padmanabhan 2001).

Fig. 1 shows the results using these late data. We obtain a characteristic 
period $\bar{P}$(t=4339 days) of 21 s; thus a pulsar born with a longer 
rotation period would remain below the bolometric luminosity limit no matter 
what its magnetic dipole moment is. The characteristic magnetic moment 
$\bar{\mu}$ (t = 4339 days) is 2.4 $\times$ 10$^{34}$ G cm$^3$ for a neutron star 
with moment of inertia 
I = 10$^{45}$ gm cm$^2$. For initial periods P$_0$ of 10, 15, or 30 ms, we 
find that two alternatives are allowed. The magnetic dipole moment is either 
less than 2.8 $\times$ 10$^{27}$, 6.4 $\times$ 10$^{27}$, or 
2.5 $\times$ 10$^{28}$ G cm$^3$, 
respectively for these initial periods, or it is greater 
than $\cong$ 2 $\bar{\mu}$ = 4.8 $\times$ 10$^{34}$ G cm$^3$. 
For a slow initial rotation rate, eg. P$_0$ = 0.3 s, we 
find $\mu < 2.5 \times$ 10$^{30}$ or $\mu > 2.4 \times$ 10$^{34}$ G cm$^3$. 
For a pulsar born with a 2 ms rotation rate, $\mu < 1.1  \times$ 10$^{26}$ or $\mu > 2.4 \times$ 10$^{34}$ G cm$^3$.

\section{Discussion}

By using the bolometric luminosity of the central remnant in SN 1987A, one can obtain 
interesting constraints on the properties of the neutron star at birth. 
The constraints on initial rotation period and on magnetic dipole moment are not 
independent of each other. To place the options in context, let us recall the 
possibilities presented by the conventional interpretation of the radio pulsar population.

Regarding the initial periods, the population of ``fresh" normal radio pulsars 
(as distinct from the ``recycled" binary and millisecond radio pulsars 
(Alpar et al. 1982, Radhakrishnan \& Srinivasan 1982) 
implies P$_0 \sim$ 10-30 ms, by extrapolation from the youngest pulsars.
An alternative suggestion, particularly for the possible pulsar in SN 1987A, 
is that it was formed by a core merger (Chen \& Colgate 1995, Middleditch et al. 2000). 
The pulsar is expected to have an initial period in the millisecond 
range, and millisecond pulsars in general can be young objects according to this scenario.

The suggestion that some pulsars are born with slow rotation rates 
to explain the distribution in the P-$\dot{P}$ diagram 
has proved difficult to check conclusively because of a 
complex of selection effects and has led to conflicting results. 
Slow initial rotation rates are not supported by the later work (Lorimer et al. 1993).  
The distribution in the P-$\dot{P}$ diagram can be obtained if pulsars 
are born with short initial periods of 10-30 ms and evolve under 
varying torques that include contributions other than the dipole 
radiation torque at some epochs in the pulsars' evolution. This 
is supported by the observation of braking indices less than the 
dipole braking index n=3 (Kaspi et al. 2001) and ages that differ 
significantly from the characteristic age P / 2$\dot{P}$ of dipole 
spindown (Gaensler \& Frail 2000). 

The dipole magnetic moments of the majority of radio pulsars are on 
the order of 10$^{30}$ G cm$^3$ (B $\sim$ 10$^{12}$ G). There is no 
evidence for decay of the magnetic moment during the normal radio 
pulsars' active lifetimes (Bailes 1989, Bhattacharya et al. 1992). 
There are a few pulsars in the tail of the distribution, with dipole 
magnetic moments in the 10$^{31}$ G cm$^3$ range (Camilo et al. 2000, 
Morris et al. 2002, McLaughlin et al. 2003) and surface dipole magnetic 
fields in the magnetar range, higher than the quantum critical 
field B$_c \equiv ({m_{e}}^2 c^3)/(e \hbar)$ = 4.4 $\times$ 10$^{13}$ G. 
PSR J1847-0130
(McLaughlin et al. 2003) has the highest dipole 
moment, $\mu$ = 9.4 $\times$ 10$^{31}$ G cm$^3$, 
corresponding to a dipole magnetic field of 9.4 $\times$ 10$^{13}$ G on 
the magnetic equator and 1.9 $\times$ 10$^{14}$ G on the poles. These 
pulsars lie in the upper right hand corner of 
the P-$\dot{P}$ diagram, among the anomalous X-ray pulsars (AXPs) 
and the soft gamma-ray (burst) repeaters (SGRs) for which the 
magnetar model has been developed (see Thompson 2000 for a review). 
While the magnetar model successfully addresses many properties of 
the SGRs and AXPs, the period clustering and the presence of radio 
pulsars with periods and inferred magnetic dipole moments similar 
to those of AXPs and SGRs has motivated another class of models 
involving fall-back disks and a combination of dipole radiation 
and disk torques (Alpar 2001, Chatterjee, Hernquist \& Narayan 2000, 
Ek\c{s}i \& Alpar 2003). In these models the magnetic moments of 
pulsars like PSR J1847-0130, and of AXPs and SGRs can be in the 
conventional 10$^{30}$ G cm$^3$ range. The absence or presence 
and mass of such disks introduces a third neutron star parameter 
at birth, in addition to P$_0$ and $\mu$. We shall assume here 
that the possible pulsar in SN 1987A has no disk or a light enough 
disk mass that its early evolution is determined by magnetic dipole 
radiation, as must be the case for the typical radio pulsars. An 
analysis of the present bolometric luminosity constraints allowing 
for fallback disks in the initial conditions will be the subject of 
separate work.
Inferred dipole magnetic moments of 10$^{27}$ - 10$^{28}$ G cm$^3$ 
are typical of the observed millisecond pulsars. 
The recycling hypothesis 
(Alpar et al 1982, Radhakrishnan \& Srinivasan 1982) is supported by the locations 
of binary and millisecond pulsars in the P-$\dot{P}$ diagram, and particularly by 
the discovery of millisecond rotation periods from low mass X-ray binaries 
(Wijnands \& van der Klis 1998), of which now five are known. These links, 
as well as the abundance of millisecond and binary pulsars in globular clusters, 
suggest that the radio pulsars with such weak inferred magnetic dipole moments 
are from an old population. The possible pulsar in SN 1987A is not expected 
to have a magnetic moment on the order of 10$^{28}$ G cm$^3$ if only old pulsars 
can have such weak fields. However, if the pulsar was born as a result of a core merger, 
allowing it a millisecond period at birth, then the magnetic field could be weak.  
Another possibility is that pulsars are born with weak initial magnetic fields, and 
the dipole magnetic fields of conventional 
$\mu \sim$ 10$^{30}$ G cm$^3$ pulsars and of magnetars are generated subsequently on 
timescales longer than the 16 yr time span of SN 1987A observations 
(see Reisenegger (2003) for a review and Michel (1994) for applications 
to the possible pulsar in SN 1987A ). 

Allowing all pulsar birth scenarios with appropriate choices of initial periods, we found that for 10, 15, and 30 ms initial periods, the upper limits 
on weak initial magnetic moments are 2.8 $\times$ 10$^{27}$, 6.4 $\times$ 10$^{27}$, or 
2.5 $\times$ 10$^{28}$ G cm$^3$. With the assumption of a slow rotator, say with an initial period of 0.3 s,  
it was found that the constraint at the weak field end was not very stringent, $\mu < 2.5 \times$ 10$^{30}$ G cm$^3$. 
If the pulsar was born from a core merger, 
with an initial period of 2 ms, then the upper limit on a weak magnetic moment is as low as $\mu < 1.1 \times$ 10$^{26}$ G cm$^3$. 

For all choices of initial period considered above, if the initial magnetic moment is not weak, and if the putative pulsar is spinning down under a constant magnetic dipole 
radiation torque, without field generation, we conclude that the 
putative pulsar has a magnetic dipole moment $\mu$ greater than  
2.4 $\times$ 10$^{34}$ G cm$^3$. If weak initial magnetic moments can be ruled out for all 
young pulsars, then we find that the possible pulsar in SN 1987A has to be a strong magnetar. 
If that is the case, then we may need to change our strategy 
for detecting the putative pulsar and look for its transient features, such as bursts observed 
from anomalous X-ray pulsars and soft gamma-ray repeaters, 
if the uniting property of all these sources is the possession of magnetar fields.

\acknowledgments
We thank Jay Gallagher, Don Cox, Chryssa Kouveliotou and Ersin G\"{o}\u{g}\"{u}\c{s} 
for useful discussions. H.\"{O}. thanks Sabanc{\i} University for hospitality during 
the preparation of the manuscript.  We thank the Turkish Academy of Sciences, and the 
High Energy Astrophysics Working Group of  T\"UBITAK for support. 
\pagebreak

\begin{figure}
\plotone{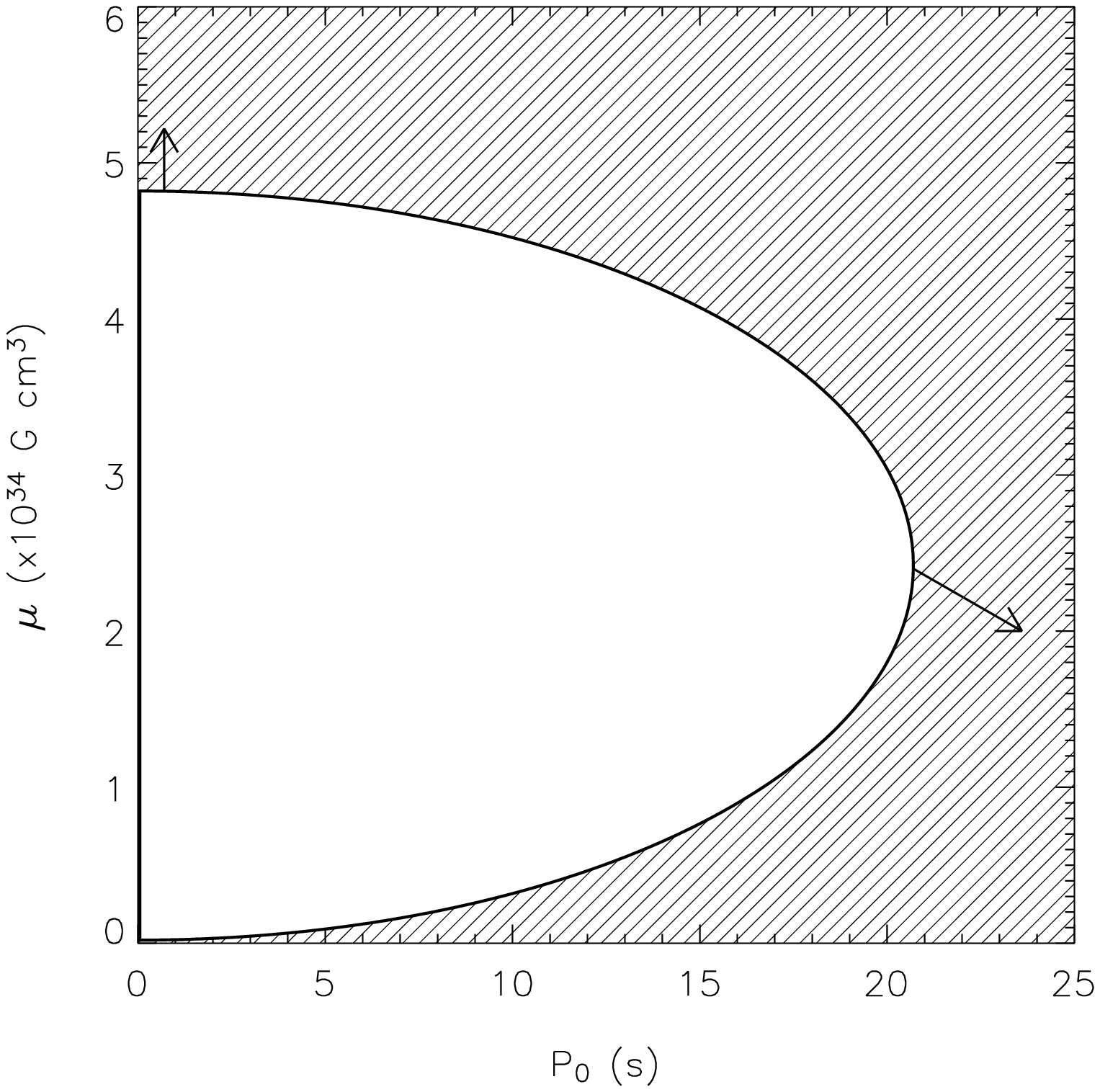}
\caption{The initial period P$_0$ and dipole magnetic moment $\mu$ of the putative pulsar allowed by the bolometric luminosity of SN 1987A must lie in the shaded region exterior to the ellipse shown. The semiminor axis $\bar{\mu}$ and the semimajor axis $\bar{P}$ are determined by L$_{bol}$(t) at each observation time t. As L$_{bol}$(t) decays exponentially, the extreme points of the ellipse move as indicated by the arrows, so that the ellipses at later times completely include the ellipses at earlier times and the tightest constraint is provided by the latest observations. The ellipse shown corresponds to the latest observations published by Soderberg et al (1999). \label{fig1}}
\end{figure}

\end{document}